\documentclass[conference,compsoc]{IEEEtran}

\ifCLASSOPTIONcompsoc
  \usepackage[nocompress]{cite}
\else
  \usepackage{cite}
\fi

\ifCLASSINFOpdf
\else
\fi

\IEEEoverridecommandlockouts

\usepackage{amsmath,amssymb,amsfonts}
\usepackage{algorithmic}
\usepackage{graphicx}
\usepackage{textcomp}
\usepackage{bmpsize}
\usepackage{xcolor}
\usepackage{lipsum}
\usepackage[colorlinks=true,urlcolor=black]{hyperref}
\usepackage{multirow}
\usepackage{booktabs}
\usepackage{array}
\usepackage{makecell}
\usepackage{threeparttable}
\usepackage{cleveref}
\usepackage{xspace}
\def\xmark{\ding{55}}
\usepackage{pifont}
\usepackage{xcolor,colortbl}
\usepackage{float}

\definecolor{LightGray}{gray}{0.98}
\definecolor{LightCyan}{rgb}{0.88,1,1}
\definecolor{Gray}{gray}{0.85}

\newcolumntype{a}{>{\columncolor{Gray}}c}
\newcolumntype{b}{>{\columncolor{LightGray}}c}

\def\BibTeX{{\rm B\kern-.05em{\sc i\kern-.025em b}\kern-.08em
    T\kern-.1667em\lower.7ex\hbox{E}\kern-.125emX}}
\begin{document}

\makeatletter
\newcommand{\newlineauthors}{%
  \end{@IEEEauthorhalign}\hfill\mbox{}\par
  \mbox{}\hfill\begin{@IEEEauthorhalign}
}
\makeatother

\title{SoK: Acoustic Side Channels} 


\author{\IEEEauthorblockN{Ping Wang}
\IEEEauthorblockA{Xidian University \\
pingwangyy@foxmail.com
}
\and
\IEEEauthorblockN{Shishir Nagaraja}
\IEEEauthorblockA{Newcastle University\\
shishir.nagaraja@ncl.ac.uk 
}
\and
\IEEEauthorblockN{Aurélien Bourquard}
\IEEEauthorblockA{
Massachusetts Institute of Technology\\
aurelien@mit.edu
}
\and
\IEEEauthorblockN{Haichang Gao }
\IEEEauthorblockA{Xidian University \\
hchgao@xidian.edu.cn
}
\newlineauthors
\IEEEauthorblockN{Jeff Yan}
\IEEEauthorblockA{University of Strathclyde\\
jeff.yan@strath.ac.uk
}
}

\maketitle

\thispagestyle{plain}

\begin{abstract}

We provide a state-of-the-art analysis of acoustic side channels, cover all the significant academic research in the area, discuss their security implications and countermeasures, and identify areas for future research. We also
make an attempt to bridge side channels and inverse
problems, two fields that appear to be completely isolated from
each other but have deep connections.
\end{abstract}


\section{Introduction}\label{sec:introduction}

Acoustic side channels (ASCs) have had a long history of interest. In the 1950's, the British intelligence spied on acoustic emanation of an Egyptian embassy's cipher machine \cite{wright1987spy}. This was a case of using sounds emitted by a Hagelin rotor machine for a side-channel attack, thereby recovering its secret key. 

The National Security Agency in the USA also had a curious and keen interest in acoustic emanation for long. It was a part of their TEMPEST program, although unsurprisingly much of the program was on leaking electromagnetic emanations.
According to the NSA's NACSIM 5000 document \cite{nsa19825000}, produced in 1982 and now partially unclassified, the TEMPEST documents NACSEM 5103, 5104 and 5105 are about acoustic emanations. But they remain classified. It was stated in \cite{nsa19825000} that `{\it Keyboards, printers, relays – these produce sound, and consequently can be sources of compromise}', but 
no further details are provided.

In the unclassified world (academia and beyond), Briol \cite{briol1991emanation} showed in 1991 that acoustic emanations of matrix printers carry, and thus leak substantial information about the printed text. Ten years later, UC Berkeley researchers Song et al. \cite{song2001timing} observed that time intervals between consecutive keystrokes leak information about the keys typed. This would make an ASC, if and only if the inter-keystroke timing is captured via acoustics. Instead, the Berkeley team exploited their neat insight for a timing side-channel attack on SSH, which relied on (network) packet timing and would give them about a factor of 50 advantage in guessing a password.
The study of keyboard emanation by Asonov and Agrawal~\cite{asonov2004keyboard} in 2004 was a landmark paper on ASCs. In the same year, Adi Shamir et al. \cite{shamir2004acoustic} announced in the rump session at Eurocrypt'04 that 
RSA decryption/signature operations running on a PC would sound differently for different secret keys. This suggested acoustic cryptanalysis become possible. It was unclear by then how to extract individual key bits from such acoustic emanations, until Shamir's team (Genkin et al. \cite{GenkinRSAAcoustic}) figured out the technical details in 2014. Since 2004, the field of ASCs started to grow rapidly, with many academic papers being published in the years to come.

Our paper represents the first (comprehensive) effort in systematising knowledge of ASCs discovered to date. We aim to make the following contributions.

 First, we will clarify some conceptual ambiguity within side-channel literature. Basic and key concepts are not defined adequately, or not at all. Consequently, the literature as a whole presents a confusing and sometimes chaotic picture. Some attacks are in fact not ASCs, but were treated as such; others are indeed ASCs but were not perceived as such. For example, does the Dolphinattack~\cite{zhang2017dolphinattack} exploit an ASC? Is Lamphone~\cite{lamphone} an ASC attack? How do ASCs and acoustic covert channels (ACCs) differ? A number of authors have presented different and even contradicting views. To tidy up things, we will introduce intuitive definitions that are simple, clear-cut and easy to operationalise. We will also introduce rigorous formal definitions, when necessary. Moreover, we will put ASCs in perspective, clarifying ASC vs ACC vs signal injection attacks, and elaborate the boundary between similar-looking but fundamentally different attacks. 
 
 Second, we will establish a taxonomy to map out, structure and evaluate the ASCs discovered to date. We will also apply a structured framework to analyse countermeasures proposed to address these ASCs.

 Third, we will perform a meta analysis of the state of the art, identifying its strengths and weaknesses. We will also offer new insights, and identify 
 future research directions.

Moreover, we make  
an attempt to bridge side channels and inverse
problems, two fields that appear to be completely isolated from
each other but have deep connections.

\section{Tidy Up the Mess}

\subsection{Ambiguity, Confusion and Possible Root Causes}
\label{sec:ambiguity}

It is not always straightforward to determine whether an attack is a side channel or not. 
Sometimes it can be tricky. Misconceptions have scattered around in the literature. 
For example,  
a well-cited paper on voice assistant security \cite{edu2020smart} mistakenly treated the Dolphinattack \cite{zhang2017dolphinattack} as a side channel, although it is  
a signal injection attack which involved with no side channel.  
Similarly, the long-range dolphin attack \cite{nsdi18} and the attack of `light commands' \cite{lightcommand} were classified as side-channel attacks in \cite{edu2020smart}. In fact, they are both not. 
On the other hand, 
some attacks (e.g. \cite{das2014you,zhou2014acoustic,zhou2018patternlistener,sonarsnoop_similar}) were indeed ACSs, but 
their authors 
did not make it explicit at all. 
More examples can go on and on. One cannot help wondering: 
what have caused such ambiguity, confusion or even mistakes? Our contemplation leads to three possible root causes as follows. 

\textbf{Root cause 1: lack of a definition of side channels that is both widely  
applicable and easy to operationalize.} 

Many papers in the literature simply used the term of `side channels' without any 
definition. This practice would work at early stages of the field, when the attacks were either a straightforward side channel or not, and many lookalike or related attacks were not invented yet. However, without a generally accepted and widely applicable definition,  it will for sure invite for ambiguity and confusion. 

On the other hand, many definitions of side channels are available in the literature, but they are different from each other, and are not very useful. Some are too narrow; perhaps more importantly, some are not {\it operational}---you cannot readily apply it to determine whether an attack is a side channel or not. We quote several definitions from the literature as follows. 

`{\it An attack enabled by leakage of information from a physical cryptosystem. Characteristics that could be exploited in a side-channel attack include timing, power consumption, and electromagnetic and acoustic emissions.}'\cite{NIST-ASC}. This NIST definition was driven by side-channel cryptanalysis, and it did not cover non-cryptanalytic side channels. It is also difficult to operationalize this definition. 

`{\it Physical side-channel attacks extract information from computing systems by measuring unintended effects of a system on its physical environment.}' Used in a recent Oakland paper \cite{Synesthesia_2019}, this definition is hard to operationalize.  

`{\it This can often be
accomplished by means of a side-channel attack, whereby an
unintended information source is leveraged.}' This definition was introduced in a recent Oakland SoK paper \cite{monaco2018sok}. It is neat, but too brief, too abstracted, and operationally not very helpful. 

`{\it 
... a side-channel attack is any attack based on information gained from the implementation of a computer system, rather than weaknesses in the implemented algorithm itself (e.g. cryptanalysis and software bugs).}' From Wikipedia, this definition is clearly driven by 
cryptanalysis and of a limited scope.

\textbf{Root cause 2: side channels and covert channels have subtle differences, and some new attack class can further complicate this subtlety.}

First, side channels and covert channels are two concepts that are related and easy-to-confuse.
For example, CovertBand \cite{ShyamnathCovertband} examined the privacy implication of tracking human movements with acoustics. It makes a clever covert channel leaking people’s privacy information, e.g. whether someone was in a room or not, or whether she was moving or standing still. But this is not a side-channel attack, as the leakage was not unintentional but on purpose. 

Second, the definitions of side channels quoted earlier ALL fail to give an angle to differentiate between side channels and covert channels. 

Third, as we will clarify later, some new class of attacks (e.g. active acoustic side channels) make it harder than before even for experts to tell whether they are a side channel or a cover channel.

\textbf{Root cause 3: The surge of similar looking but different acoustic attacks has further complicated the conceptual ambiguity and confusion in the field.}

Acoustic security has expanded rapidly and substantially in the recent years. 
Acoustic attacks such as 
the Dolphinattack \cite{zhang2017dolphinattack}, the long-range dolphin attack \cite{nsdi18} and the `light commands' attack \cite{lightcommand}, discussed earlier, represent only a single class of sources for confusions. There are more.

Another set of acoustic attacks eavesdrop and recover human speech by picking up vibrations via motion sensors, cameras, laser or lidar, e.g. \cite{michalevsky2014gyrophone,oakland18,lamphone,monaco2018sok,han2017pitchln,roy2016listening}. They represent another class of confusion sources. These attacks involved with side channels, but not necessary acoustic ones.
For example, a gyroscope's reading is sensitive to sound vibrations, and Stanford researchers Michalevsky et al.~\cite{michalevsky2014gyrophone} used it to recover human speech. This is a side-channel attack, but not an acoustic one. Only when the vibration frequency is in a certain range (20$\sim$20KHz), the signal is acoustic. The Lamphone attack~\cite{lamphone} recovers human speech by measuring vibrations of a light bulb caused by acoustic waves. However, it exploits an optical side channel, rather than an acoustic one, to recover the sound.

\subsection{Our Definitions}
A key aspect of side channels is unintended communication. Acoustic energy is present as wave energy within an air medium, or as vibrations within solid media. Formally, a {\it \textbf{side channel is a communication channel which allows one-way information transfer from the targeted system to the attacker. A side channel is defined as a functional mapping $S: I \times M \mapsto O$, where $I$ is the valid input system inputs, $M$ is the attacker's influence on $S$, and $O$ are the observations made by an attacker monitoring the channel. The attacker's goal is to {\em infer} $I$ from observations $O$.}} In an ASC, the system leaks information acoustically, i.e. observations $O$ are made on an acoustic medium regardless of any adversarial influence $M$, the influence mechanism or the influence medium. Not all side channels involve adversarial influence, in which case $S$ is termed as a \emph{passive} side channel (when $|M|=0$). However, in the presence of adversarial influence $|M| > 0$, $S$ is termed as an \emph{active} side channel. Note that key distinguishing characteristic of an ASC is that the attacker can only observe the victim over the acoustic channel. The scenario where adversarial influence is over an acoustic channel whilst observations are made on a non-acoustic channel is not an ASC. When no information is leaked, i.e. $O$ is $NULL$, then there is no side channel in existence, even if the adversary is able to influence the system. This is the dual of the active side channel and is termed as a {\em signal-injection attack}.

To address the ambiguity and confusion in the field (see Section~\ref{sec:ambiguity}), we have developed definitions by first organising the research landscape on the basis of attacker and defender capabilities or {\em threat-models} (See Figure~\ref{fig:taxonomy}). Threat models can be classified based on two factors namely the physical channel the attacker can access (eg. acoustic) and transmissions (receive-only (Rx), transmit-only (Tx), or send-and-receive (Rx,Tx)). An attacker is denoted as $M^C_F$ where $C \in U$ is the set of channels the attacker can access out of the universal set of possibilities $U$, and $F$ is a subset of $\{tx,rx\}$. The target $T^{C'}_{F'}$ is similarly defined in terms of channels accessed  $C \in U$ and channel functions $F' \in \{tx,rx\}$. The combinations of possible attacker and target profiles define the threat landscape.


\begin{figure}[htbp]
    \centerline{\includegraphics[width=1\columnwidth]{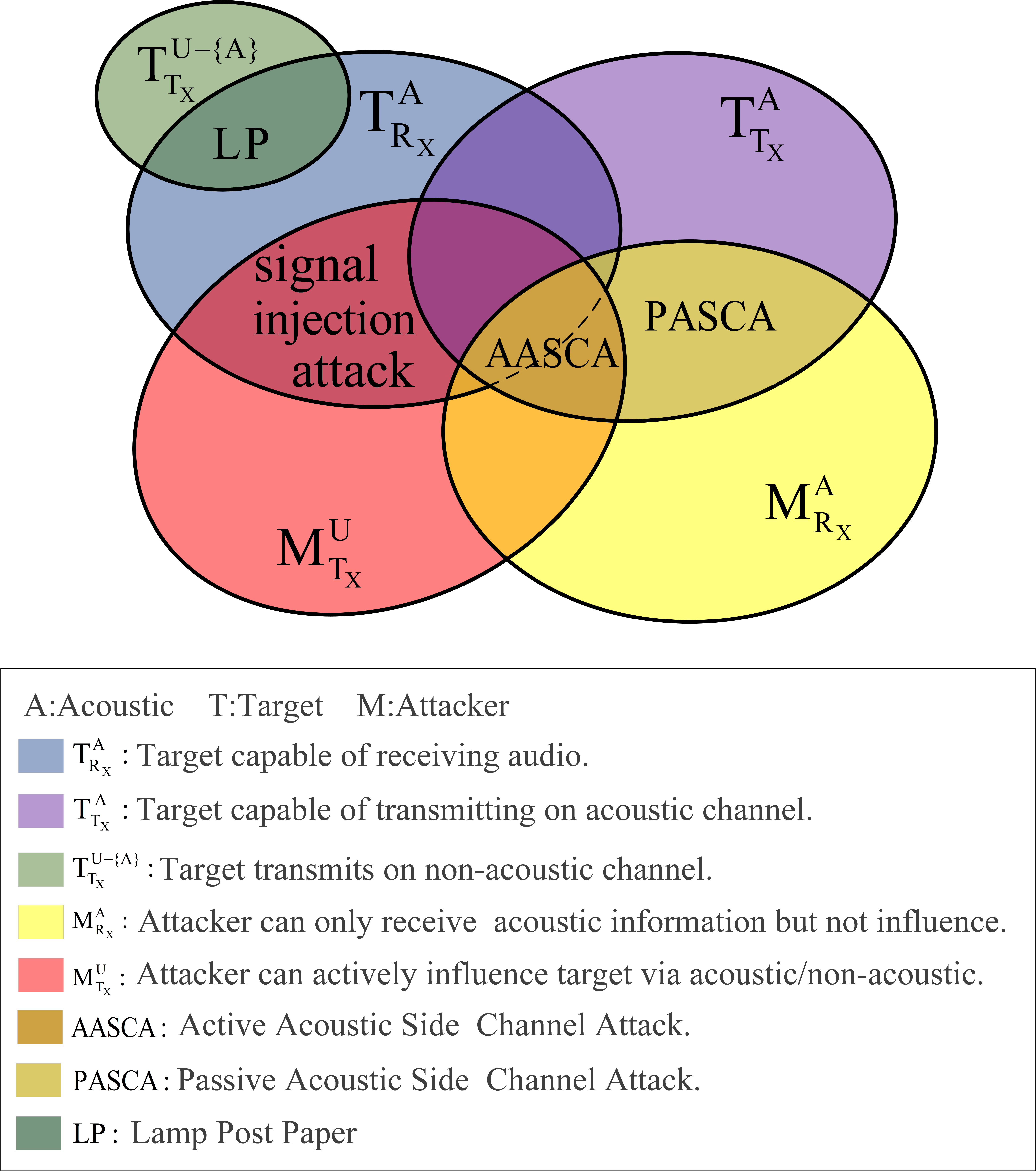}}
    \caption{Venn diagram mapping threat models to side channel taxonomy}
    \label{fig:taxonomy}
\end{figure}

Side channels can be exploited either for {\em offensive}~\cite{asonov2004keyboard,das2014you,sonarsnoop,voiploc:wisec21,zhou2018patternlistener} or {\em defensive} purposes~\cite{belikovetsky2018digital,usenixsec17,panda2020behavioral}. When used for attack, one of the channel endpoints, typically the source of information leakage is the defender, while the sink is the attacker. Side channels as defenses are possible where the attacker is replaced by a defender.

A {\em Passive Acoustic Side-Channel Attack (PASCA)} is characterised by an attacker $M^A_{Rx}$ who can only monitor the acoustic channel while the victim $T^A_{Tx}$ only transmits. A PASCA is a receiver-only channel for the attacker and a transmission-only channel for the victim. Therefore, the victim cannot be influenced by the attacker. 
Figure~\ref{fig:taxonomy} visualises the landscape and suggests boundaries between the various notions of abusing unintended communication channels on the basis of threat models.  An {\em Active Acoustic Side-Channel Attack (AASCA)} is characterised by a victim  $T^{A}_{Rx,Tx}$, who is unintentionally transmitting information over an acoustic channel and an attacker who can make observations. Additionally, the attacker can also influence the victim via another channel which can either be acoustic or non-acoustic to induce a change in leakage behaviour i.e. change the rate of leakage or what is leaked through the acoustic channel. The crucial difference from PASCA is that the attacker influences the victim. Note that influence can be either via non-acoustic or acoustic means, as long as the target leaks information acoustically, we have an ASC attack.

AASCA is relatively more powerful than its PASCA counterpart. The ability to influence a victim means that an attacker can induce leakage to optimise inference. On the other hand, a PASCA is stealthier since the attacker is not transmitting any information that can be used by the defender to detect and isolate the attacker. For example, the transmission of ultrasound or mechanical vibration by the attacker may be observed by the defender, thus making active attacks relatively detectable. A successful side-channel attacker must therefore draw a balance between the active and stealth components of their attack.


Different from side channels but related, an {\em Acoustic Covert Channel Attack} (ACCA) involves two or more attackers who are communicating over a channel that is unintentionally present i.e. the endpoints are intentional but the channel is unintentionally present. Thus a covert channel differs from side channels primarily in the functional mapping $S:I \times M \mapsto O$ in the following important way: in a side channel the function $S$ is defined by the victim and the attacker has no control over it. In a covert channel, both ends are under attacker control, therefore the attacker can optimally define and implement $S$ such that hidden information $I$ can be readily inferred from observations $O$. In a covert channel, the leak is deliberate as the attacker controls both channel endpoints, whereas in a side channel the attacker does not control the source endpoint. Covert channels were first described by Lampson~\cite{lampson1973note}.

Due to their similarity, side channels and covert channels are often confused for one another. As one example, SonarSnoop~\cite{sonarsnoop} is a side channel rather than a covert channel attack. In SonarSnoop, speakers are used to emit human inaudible acoustic signals and the echo is recorded via microphones, turning the acoustic system of a smartphone into a sonar system. The echo signal from a user's finger movements can be inferred to steal Android phone unlock patterns. In this attack, indeed acoustic signals were intentionally induced, but the researchers measured only echoes from finger movements, which did not  deliberately leak information i.e. source endpoint is not under attacker control. As the transmission was accidental, SonarSnoop is a side-channel attack rather than a covert-channel attack.

Another comparative point is that in the case of side-channel-as-defense, the defender has no control over source behaviour. For instance, they cannot make changes to the keyboard in order to enable the generation of optimal acoustic signatures. However, that changes when we consider an {\em Acoustic Covert-Channel-as-Defense} (ACCaD). An example ACCaD would be the use of an unintentional communication channel between systems at the same security level perhaps to fulfill a monitoring function. To the best of our knowledge, no ACCaD has been proposed thus far. 

Often, a direct measurement of the output from a side channel does not immediately give the information leaked via the channel. And the channel output is more like meta data, from which attackers deduce the leaked information in a sensible way to complete their attacks. An exception is transient execution attacks such as Meltdown~\cite{lipp2020meltdown} and Spectre~\cite{kocher2020spectre}, which are side channels that leak actual data, instead of meta data. In contrast, traditional micro-architectural side-channel attacks leak only meta data, such as memory access patterns.


\section{Acoustic Side Channels: A Taxonomy}
\label{sec:attacks}

To classify ASCs, we consider leaking devices, the leaked signals, the media via which the leakage occurs, as well as the information leaked. We also consider various features of each ASC, such as whether it is an active or passive attack, whether it is used for offensive or defensive purposes, the attacker's distance, and the signal properties. We propose 
a taxonomy as in \Cref{tab1}, whereas its high-level structure is
illustrated in Figure~\ref{fig:ASCtaxonomy}.   
Our taxonomy categories highlight the most interesting ASC characteristics. 

\begin{figure}[htbp]
    \centerline{\includegraphics[width=1\columnwidth]{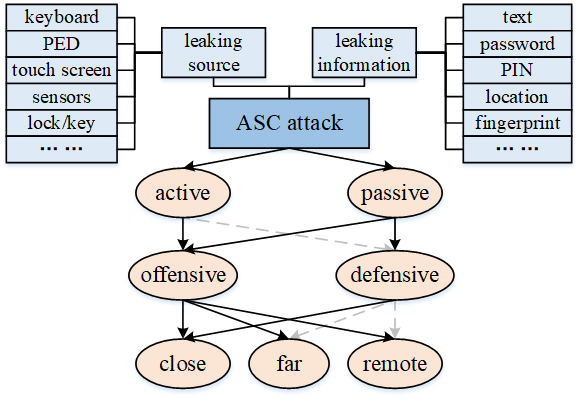}}
    \caption{The structure of our ASC taxonomy: a high-level view (dash lines  represent possible combinations but no such papers published yet)}
    \label{fig:ASCtaxonomy}
\end{figure}

\begin{table*}[htbp]
\linespread{0.6}
\setlength\tabcolsep{3pt}
    \footnotesize
    \centering
    \vspace{-0.7cm}
    \resizebox{\textwidth}{!}{
    \begin{threeparttable}[b]
    \caption{Acoustic side channels: a taxonomy}
    \label{tab1}
    
    \begin{tabular}{cc|cc|cccccc}
    \toprule
   \multicolumn{2}{c}{} &  \multicolumn{2}{|c|}{\textbf{Leaking}} &  \multicolumn{5}{c}{} &   \multicolumn{1}{c}{\textbf{Sampling}} \\
   \textbf{Categories} & \textbf{Ref.} & \textbf{Source} & \textbf{Information} & \textbf{Audible} & \textbf{Purpose} & \textbf{Active} & \textbf{Intrusive}  & \textbf{Proximity}\tnote{1}  & \textbf{frequency} \\ \midrule

    \multirowcell{9}{Keyboard \\ emanation} & \makecell{Asonov'04~\cite{asonov2004keyboard} } & \makecell{Physical keyboard} & Typed text & \textcolor{blue}{\checkmark}& offensive& \textcolor{red}{\xmark} & \textcolor{red}{\xmark} & close, far & \makecell{44.1KHz} \\  \cmidrule(l){2-10}

    & \makecell{Zhuang'05~\cite{zhuang2005keyboard}} & \makecell{Physical keyboard} & Typed text & \textcolor{blue}{\checkmark}& offensive& \textcolor{red}{\xmark} & \textcolor{red}{\xmark} & close, far & \makecell{0.4$\sim$12KHz} \\  \cmidrule(l){2-10}
    
    &\makecell{Berger'06~\cite{Berger:2006:DAU:1180405.1180436} \\  Zhu'14~\cite{Zhu:2014:CAU:2660267.2660296} \\ Helavi'15~\cite{ijis15} \\ Slater'19~\cite{acsac19}} & \makecell{Physical keyboard} & Typed text & \textcolor{blue}{\checkmark}& offensive& \textcolor{red}{\xmark} & \textcolor{red}{\xmark} & close &  44.1KHz\\    \cmidrule(l){2-10}
    
    & \makecell{Liu'15~\cite{Liu:2015:SKM:2789168.2790122}} & \makecell{Physical Keyboard} & Typed text & \textcolor{blue}{\checkmark}& offensive& \textcolor{red}{\xmark} & \textcolor{red}{\xmark} & close & \makecell{48KHz\\192KHz}\\  
    \cmidrule(l){2-10}
    
    & \makecell{Martinasek'15~\cite{martinasek2015acoustic}} & \makecell{Physical keyboard} & Typed text & \textcolor{blue}{\checkmark}& offensive& \textcolor{red}{\xmark} & \textcolor{red}{\xmark} & close & 48KHz\\  
    \cmidrule(l){2-10}
    
   & \makecell{Ranade'09~\cite{ranade2009acoustic}  }& PED & Key taps & \textcolor{blue}{\checkmark}& offensive& \textcolor{red}{\xmark} & \textcolor{red}{\xmark} & close & 44.1KHz \\  
    \cmidrule(l){2-10}
    
   & \makecell{Cardaioli'20~\cite{cardaioli2020your} }& PED & Key taps & \textcolor{blue}{\checkmark}& offensive& \textcolor{red}{\xmark} & \textcolor{red}{\xmark} & close & 48KHz \\  
    \cmidrule(l){2-10}
    
    & \makecell{Panda'20~\cite{panda2020behavioral}} & PED & \makecell {Key taps \& \\ User identity } & \textcolor{blue}{\checkmark}&  \makecell{ offensive \& \\  defensive }  &\textcolor{red}{\xmark}& \textcolor{red}{\xmark} & close  & 0.04$\sim$20KHz\\ \cmidrule(l){2-10}
    
    & Enigma'15~\cite{ToreiniAcousticEnigma} & \makecell{Enigma keyboard} & Key taps & \textcolor{blue}{\checkmark}& offensive& \textcolor{red}{\xmark} & \textcolor{red}{\xmark} & close & 44.1KHz\\  
    \midrule
    
     \multirowcell{4}{Acoustic \\ finger-tapping \\ emissions} &   \makecell{Narain'14~\cite{SashankMicandGyro}} & \makecell{Touch screen} & Typed text & \textcolor{red}{\xmark}& offensive& \textcolor{red}{\xmark} &\textcolor{blue}{\checkmark} & close & 48KHz \\ \cmidrule(l){2-10}
     
    &  \makecell{PIN Skimmer'13~\cite{Simon:2013:PSI:2516760.2516770}} & \makecell{Touch screen} & Typed text & \textcolor{red}{\xmark}& offensive& \textcolor{red}{\xmark} &\textcolor{blue}{\checkmark} & close & 16KHz  \\  \cmidrule(l){2-10}
    
     & \makecell{Shumailov'19~\cite{shumailov2019hearing}} & \makecell{Touch screen} & Typed text & \textcolor{red}{\xmark}& offensive& \textcolor{red}{\xmark} &\textcolor{blue}{\checkmark} & close & 44.1KHz\\ \cmidrule(l){2-10}
     
     & Zarandy'20~\cite{zarandy2020hey} & Touch screen & Typed text & \textcolor{red}{\xmark} & offensive &\textcolor{red}{\xmark} &\textcolor{blue}{\checkmark}  & close  & 48kHz\\
    \midrule
    
     \multirowcell{3}{Acoustic \\ motion \\ detection} &   SonarSnoop'18~\cite{sonarsnoop} & \makecell{Human-Computer \\ Interaction} & \makecell{Gesture \\password} & \textcolor{red}{\xmark} & offensive  & \textcolor{blue}{\checkmark} &\textcolor{blue}{\checkmark} & close & 48KHz \\ \cmidrule(l){2-10}

     & KeyListener'19~\cite{lu2019keylistener} & \makecell{Human-Computer\\ Interaction} & Typed text & \textcolor{red}{\xmark} & offensive &\textcolor{blue}{\checkmark} & \textcolor{blue}{\checkmark}    & close  & 20kHz\\ \cmidrule(l){2-10}
         
     & \makecell{PatternListener'18~\cite{zhou2018patternlistener} \\PatternListener+'19~\cite{sonarsnoop_similar}} & \makecell{Human-Computer \\ Interaction} & \makecell{Gesture \\password} & \textcolor{red}{\xmark} & offensive  & \textcolor{blue}{\checkmark} &\textcolor{blue}{\checkmark} & remote & 48KHz \\
     \midrule

    \multirowcell{3}{VoIP \\ hitchhiking \\ ASC} & \makecell{Skype \& Type'17~\cite{DBLP:journals/corr/CompagnoCLT16} }  & Keyboard & Key taps & \textcolor{blue}{\checkmark} & offensive & \textcolor{red}{\xmark} & \textcolor{red}{\xmark} & remote & 44.1KHz \\\cmidrule(l){2-10}
    
    & Anand'18~\cite{codaspy18} & Keyboard & Key taps & \textcolor{blue}{\checkmark}& \makecell{offensive} & \textcolor{red}{\xmark} & \makecell{\textcolor{blue}{\checkmark}} & close, remote & 44.1KHz\\ \cmidrule(l){2-10}
    
    & \makecell{LendMeYourEar'22\\ ~\cite{genkin2022lend} } & EM fields (via acoustics) &  \makecell{Computation \\ dependent \\ leakage} & \textcolor{blue}{\checkmark}& \makecell{offensive } & \textcolor{red}{\xmark} & \textcolor{red}{\xmark} & remote & 48KHz
    \\ \midrule
 
   \multirowcell{2}{Physical \\ location \\ fingerprinting} & Jeon'18~\cite{www18} & \makecell{Electricity \\network} & \makecell{Physical \\location} & \textcolor{blue}{\checkmark}& offensive& \textcolor{red}{\xmark} & \textcolor{red}{\xmark} & remote & 1KHz \\
    \cmidrule(l){2-10}
    
    & VoIPLoc'21~\cite{voiploc:wisec21} & \makecell{Rooms} & \makecell{Physical \\ location} & \textcolor{blue}{\checkmark}& offensive & \textcolor{red}{\xmark} & \textcolor{red}{\xmark} & remote & 44.1KHz \\ \midrule

    \multirowcell{3}{Acoustic \\ device \\ fingerprinting} & Das'14~\cite{das2014you} & \makecell{Internal sensors} & Device ID & \textcolor{blue}{\checkmark} &	offensive & \textcolor{red}{\xmark} & \textcolor{blue}{\checkmark} & close, far & \makecell{8KHz \\ 22.05KHz \\ 44.1KHz} \\ 
    \cmidrule(l){2-10}
    
    & Zhou'14~\cite{zhou2014acoustic} & \makecell{Internal sensors} & Device ID & \textcolor{red}{\xmark} & offensive &\textcolor{red}{\xmark} &\textcolor{blue}{\checkmark}    & close, far  & 44.1KHz\\
    \cmidrule(l){2-10}
    
    & Kotropoulos'14~\cite{kotropoulos2014mobile} & \makecell{Internal sensors} & \makecell{Phone module} & \textcolor{blue}{\checkmark}  & offensive & \textcolor{red}{\xmark} & \textcolor{red}{\xmark} & close & 16KHz \\
    \midrule

    \multirowcell{9}{ASC \\ based on \\ Device Hum} & \makecell{Briol'91~\cite{briol1991emanation} \\ Backes'10~\cite{BackesAcousticPrinters}} & \makecell{Dot-matrix printer} & Printed text & \textcolor{blue}{\checkmark}& offensive& \textcolor{red}{\xmark} & \textcolor{red}{\xmark} & close & 96KHz\\
    \cmidrule(l){2-10}

    &   \makecell{Hojjati'16~\cite{HojjatiSideChannelFactory}} & \makecell{3D printer \& CNC mill} & \makecell{Proprietary \\IPR info}  & \textcolor{blue}{\checkmark}& offensive& \textcolor{red}{\xmark} & \textcolor{red}{\xmark} & close & 44.1KHz \\ \cmidrule(l){2-10}

    &   \makecell{Song'16~\cite{ccs16}} & \makecell{3D printer} & \makecell{Proprietary \\IPR info}  & \textcolor{blue}{\checkmark}& offensive& \textcolor{red}{\xmark} & \textcolor{red}{\xmark} & close & 44.1KHz \\ \cmidrule(l){2-10}

    &   \makecell{Faruque'16~\cite{FaruqueAcoustic3DPrinter} \\ Chhetri'18~\cite{ChhetriConfidentiality}} & 3D printer & \makecell{Proprietary \\IPR info}  & \textcolor{blue}{\checkmark} & offensive& \textcolor{red}{\xmark} & \textcolor{red}{\xmark} & close & 96KHz \\ \cmidrule(l){2-10}
    
    & \makecell{KCAD'16~\cite{chhetri2016kcad}} & 3D printer & Control signals & \makecell{\textcolor{blue}{\checkmark}} & defensive & \textcolor{red}{\xmark}& \textcolor{red}{\xmark} & close  & $>$40KHz\\ \cmidrule(l){2-10}

    & \makecell{Bayens'17~\cite{usenixsec17}} & 3D printer & Fill pattern & \textcolor{blue}{\checkmark} & defensive &\textcolor{red}{\xmark}&\textcolor{red}{\xmark} & close  & 44.1KHz\\ \cmidrule(l){2-10}

    & \makecell{Belikovetsky'19~\cite{belikovetsky2018digital}} & 3D printer & Audio fingerprint & \textcolor{blue}{\checkmark}& defensive &\textcolor{red}{\xmark}&\textcolor{red}{\xmark} & close  & 44.1KHz\\ 
    \cmidrule(l){2-10}

    & Synesthesia'19~\cite{Synesthesia_2019} & \makecell{LCD monitor \\ (power bank)} & \makecell{Display \\contents} & \textcolor{red}{\xmark} & offensive& \textcolor{red}{\xmark} & \textcolor{red}{\xmark} & \makecell{close, far, \\ remote} & \makecell{40KHz \\ 192KHz} \\
    \cmidrule(l){2-10}
    
    & \makecell{Islam'18~\cite{macs18}} & \makecell{Cooling fan} & \makecell{Electrical \\load} & \textcolor{blue}{\checkmark}& offensive& \textcolor{red}{\xmark} & \textcolor{red}{\xmark} & close & 8KHz \\
    \midrule

    \multirowcell{2}{Physical-key \\ leakage} & \makecell{SpiKey'20~\cite{ramesh2020listen} } & \makecell{Mechanical \\lock and key} & Physical key & \textcolor{blue}{\checkmark} & offensive& \textcolor{red}{\xmark} & \textcolor{red}{\xmark} & close & \makecell{44.1KHz} \\  \cmidrule(l){2-10}
    
     & \makecell{ Keynergy'21~\cite{usenixsec21}} & \makecell{Mechanical \\lock and key} & Physical key & \textcolor{blue}{\checkmark} & offensive& \textcolor{red}{\xmark} & \textcolor{red}{\xmark} & close & \makecell{44.1KHz \\ 192KHz} \\ \midrule

     \makecell{Acousitc \\ cryptanalysis} & \makecell{Genkin'14~\cite{GenkinRSAAcoustic} \\ Genkin'17~\cite{Genkin-2017}} & \makecell{Motherboard} & Crypto keys & \textcolor{red}{\xmark}& offensive& \textcolor{red}{\xmark} &\textcolor{blue}{\checkmark} & close & \makecell{21, 40, 48,\\200, 350KHz} \\ \midrule

    \makecell{DNA synthesis} & Oligo-Snoop'19~\cite{ndss19} & \makecell{DNA \\synthesizers} & \makecell{DNA \\sequence} & \makecell{\textcolor{blue}{\checkmark} }& offensive& \textcolor{red}{\xmark} & \textcolor{red}{\xmark} & close & 48KHz \\

    \bottomrule
    \end{tabular}
         \begin{tablenotes}
     \item[1] 
The proximity between the attacker and the target. Close:  the attacker is physically near the target 
(up to 3 meters). 
Far:  
typically 10 to 100 meters. 
Remote: 
the attacker can only access the target remotely, usually through a network connection.

     \end{tablenotes}
  \end{threeparttable}}
  \vspace{-0.7cm}
\end{table*}

\subsection{Keyboard Emanation} 

Asonov and Agrawal~\cite{asonov2004keyboard} was the first to observe that each physical key has a unique acoustic (sound) signature as a fundamental property of keyboard design.  
Their main insight was that the physical plate beneath the keys causes each key to produce a different sound (frequency) depending on its location on the plate (similar to hitting a drum at different locations) thus these keystroke sounds can be used to steal what is being entered.  
Zhuang et al.~\cite{zhuang2005keyboard} combined per-key acoustic fingerprints with a language model in an unsupervised learning setting (KMeans+HMM) improving inference efficiency from 52\% to 67\%. Berger et al.~\cite{Berger:2006:DAU:1180405.1180436} introduced a comprehensive language model via a password dictionary.

An alternate to acoustic frequency spectrum is to leverage signal timing. Zhu et al.~\cite{Zhu:2014:CAU:2660267.2660296} observed that the relative time-of-arrival of an acoustic signal is dependent on the distance between the sensor and the originating keypress measured as the time-difference-of-arrival (TDoA) at attacker microphones placed 1m apart. Reported inference accuracy is 72\%.

Combining both signal timing and acoustic features, Liu et al.~\cite{Liu:2015:SKM:2789168.2790122}, report a recovery rate of 94\% of keystrokes. Their main insight was that combining signal warfare (TDoA) techniques with the frequency spectrum (MFCC) effectively replaced the benefits accorded by a language model, and simply running K-Means over the fingerprint vector was enough to cluster them by the key. This is significant since security practices around password construction may not permit content that is compatible with a language model.

Halevi et al.~\cite{ijis15} evaluated the impact of typing styles in key recovery rates. They observed that while keys have unique sound signatures, touch typing significantly reduces the signal-to-noise ratio reducing recovery rates to 56\% in the supervised case. They also found a significant decrease in key recovery rates when training and testing writing styles differ. The extent to which this applies to the unsupervised learning approaches above is unknown. Martinasek et al. \cite{martinasek2015acoustic} and Slater et al.~\cite{acsac19} utilized neural networks to complete classification and Slater et al. found that deep learning approaches are well suited to the task of key recovery in noisy environments.

Specialist keyboards such as Pin Entry Devices (PEDs) and ATM/PoS keypads are equally vulnerable to key transcription attacks via sound side-channels and the attacks leverage the sound produced by a keypress on ATM keypads~\cite{ranade2009acoustic} and Enigma keyboards~\cite{ToreiniAcousticEnigma}. Cardaioli et al.~\cite{cardaioli2020your} found that using inter-key delays extracted from signal arrival information works well too. This is an important improvement over Asonov's sound-of-the-key approach, since it only uses signal timing information via a single sensor (as opposed to the multi-sensor TDoA approach of Zhu et al.~\cite{Zhu:2014:CAU:2660267.2660296}). Panda et al.~\cite{panda2020behavioral} also recovered PIN keys from the keypress acoustic emanation, but they used the interval between two keystrokes as the main feature. In addition to exploiting this ASC for offensive purposes, the researchers in~\cite{panda2020behavioral} also explored it for defensive purposes. Namely, the keystroke dynamics emitted via acoustics could work as behavioural biometrics for each user, offering additional protection for their PINs in theory.

\subsection{Acoustic Finger-tapping Emissions} 
This category of attacks targets touchscreen keyboards on smartphones and tablets, instead of physical keyboards. When a user taps the screen, a fixed glass plate, with a finger, the tap generates a sound wave that propagates on the screen surface and in the air. Although signal strength is weaker than keystrokes from physical keyboards, it is well above the noise floor.

Early efforts were multi-modal---they combined acoustic information with other sources to isolate keypresses. Narain et al.~\cite{SashankMicandGyro} proposed a passive attack method to infer the text content created by taps on a touchscreen keyboard by using a Trojan application to capture sensed data from stereoscopic microphones and gyroscope. Simon et al.~\cite{Simon:2013:PSI:2516760.2516770} developed PIN Skimmer which combines device microphones to detect touch events and device orientation information from the video camera inputs, to estimate the position of the tapped number.

The first to propose a fully acoustic passive ASC attack was Shumailov et al.~\cite{shumailov2019hearing} on touchscreen keyboards. They observed that acoustic waves passing through the glass bounce off the screen sides creating unique acoustic patterns observable from the internal microphones. Authors record the audio through the built-in microphones and demonstrate that simple TDoA allows the attacker to decipher PIN rows, while more complex machine learning models can use acoustic information to recover the actual PIN code, as well as, the text typed in.

Building on findings of~\cite{shumailov2019hearing}, Zarandy et al.~\cite{zarandy2020hey} observed that voice assistants such as Amazon Alexa and Google Home can be abused by an attacker to echolocate the sounds of a key tap on a different device. 
The authors demonstrate that it is possible to perform the attack up to half a meter away from the voice assistant.

\subsection{Acoustic Motion Detection}
An {\em active} attacker can 
exploit system behaviour by introducing a {\em new side-channel}. SonarSnoop~\cite{sonarsnoop} is the first such ASC attack for detecting finger motion. The attacker deploys malware on a victim's smartphone to generate ultra-sound chirps. By analysing echoes (chirp reflection), the dynamic motion of the fingers can be reconstructed in a fine-grained resolution to support recovery of pattern passwords. In this attack, the active component is the introduction of a stealthy sound-field outside human-audible range. The attacker exploits the property that the victim unintentionally modulates the attacker signal with confidential information. The unintentional transmission is a key characteristic of a side-channel. Zhou et al~\cite{zhou2018patternlistener,sonarsnoop_similar} explored a similar approach to recover gesture passwords.
Acoustic motion detection can also be used to localise  
virtual keyboard inputs. In 2019, KeyListener~\cite{lu2019keylistener}  
developed an active ASC attack that leveraged the change in Doppler effects due to finger movement within an induced sound field, to isolate touchscreen taps. All three works are active ASC as they require an active agent (malware or external device) to induce the sound field.

 \subsection{VoIP Hitchhiking ASC}
It is natural to explore whether side channels can span (hitch-hike over) Voice over Internet Protocol (VoIP) sessions. Theoretically, this should be possible as human-voice frequency (20-20KHz) overlaps with keyboard sound frequency range (2-4KHz). Compagno et al.~\cite{DBLP:journals/corr/CompagnoCLT16} confirm this via real-world experiments over the Skype network (Opus Codec) as long as the bandwidth is more than 20bps. The technical mechanism is largely based on the same attack components as prior art (MFCC-based acoustic signature features mated with a supervised learning inference mechanism). 
Anand et al.~\cite{codaspy18} confirm that keypads and ATM PEDs are equally vulnerable to key transcription side-channel attacks over VoIP sessions as they are close-proximity attacks. This means that scammers who get victims to hand over account information and then persuade them to walk over to an ATM to `check balance' whilst on a call to the scammer, may steal their victim's PIN as well as their account information.

More recently, Genkin et al.’s~\cite{genkin2022lend} observed that the built-in microphones of PCs can inadvertently capture computation-dependent leakage with electromagnetic (EM) fields within the computer even at a remote distance. It is possible because CPU computation leaks through audio signals. They demonstrated the efficacy by exploiting the leakage to perform attacks in three different scenarios---website identification, cryptographic key recovery, and multiplayer games cheating, via remote VoIP communication.

 \subsection{Physical-location Fingerprinting}
When using VoIP to communicate, the created audios and data streams always include electrical network frequency (ENF) signals and other acoustic-reflection signals except for audible sounds. These signals always have specific characteristics and some important information, such as time and location. Therefore, it is possible to use those signals as signatures for location inference. 
Jeon et al.~\cite{www18} proposed an attack to identify the physical location of where a target video or sound was recorded or streamed from. 
To achieve the attack, they first created a reference map of ENF signals extracted from the multimedia streaming data from a victim’s device via the microphone and then extracted the location information from the map by a two-step estimation. This work belongs to a passive way and is considered an ASC attack because all the targeted information is essentially leaked from the acoustic signals of multimedia streaming data. Different from 
those that require installing a specific malicious application on a victim’s device, this attack can be performed with existing VoIP applications or online streaming services, which means the only data needed is a target multimedia file and it is non-intrusive. 

Nagaraja et al.~\cite{voiploc:wisec21} proposed a passive attack (location fingerprinting technique) for a location inference on VoIP calls via ASCs, called VoIPLoc. Specifically, it exploited the acoustic-reflection characteristics of the physical space of a VoIP user. Using the speaker voice as the impulse signal, it extracted signals and then utilized a multi-layer classifier to map the fingerprint to a location. 

\subsection{Acoustic Device Fingerprinting}
Microphones and speakers can be fingerprinted by variations in sensing and actuation respectively, introduced by variations in their physical properties. Das et al.~\cite{das2014you} note that variations in the chemical compositions of diaphragm material, aging-related changes in the mount point, the glue used, wear-and-tear in manufacturing machines, humidity, and temperature levels during manufacturing all play a role in ensuring that no two microphones or speakers come off the assembly line working identically. Given an audio sample, they were able to trace 98\% of the samples to the sensing device by using MFCC features of recorded audio. Both Zhou et al.~\cite{zhou2014acoustic} and Kotropoulos et al.~\cite{kotropoulos2014mobile} independently discovered the same phenomena and devised a speaker fingerprinting method based on high-frequency power spectrum. In summary, manufacturing imperfections have been successfully exploited to attribute audio recordings to specific devices.

\subsection{ASC based on Device Hum}

{\bf Printer hum:}
Often, electro-mechanical devices with moving physical parts are vulnerable to ASCs. Moving mechanical parts create vibrations that leak into the surroundings either as sound or as acoustic vibrations through the body of the device. In many cases, the movement of the mechanical components such as motors, fans, base plates, pins, and drums, is a function of user input leading to information leakage through acoustic channels. 
Briol~\cite{briol1991emanation} was the first to report an ASC in dot-matrix printers. Dot-matrix printers use multiple rows of needles. When printing a character, a subset of needles strike the paper surface mounted on a backing plate, a mechanical action that generates a sound wave. It turns out that printed characters generate a unique sound for each character printed (just as keyboards). It is therefore natural to expect that the approach and techniques developed for key transcription attacks are applicable to printer inference attacks. Backes et al.~\cite{BackesAcousticPrinters} confirm this---recording the sound from a microphone close enough to the printer, and passing it through a standard pipeline of basic signal processing to extract the short-term power-spectrum features (MFCC) in the relevant frequency band ($>20$KHz). The main difference with keyboards, is the characters are printed at a higher rate than human keypresses. Due to this, acoustics of keys get mixed up due to time-overlapping signals. Interestingly, the sound of printers is above 20KHz band whereas keyboards emit sound at 2$\sim$4KHz band. This means key transcription and printer inference do not interfere with each other, and can be executed simultaneously, if required. 
In comparison with key-transcription attacks, printer information leakage is relatively less developed. We know of no works that apply time-difference-of-arrival of printer sound, learning-based inference, and signal-timing information (inter-character delay period). The application of these ideas may improve the state-of-the-art in printer transcription attacks, especially the issue of separating overlapped signals.

{\bf 3D printer hum:} Different from toner-based printers, 3D printers use a motorised filament extruder which deposits layers of material via an extrusion arm, whose location is controlled by multiple stepper motors to precisely control where filament is delivered on a base plate. The amount of current supplied to the various motors depends on the (confidential) printer input. Fundamentally, motors emit sound waves as a direct result of the current applied~\cite{175275},
arising first from {\em magnetostriction}: change in material dimensions in proportion to passing current in fixed electromagnets in the motor; {\em electrostriction}: change in dimensions of the conducting coil within the motor in proportion to current passing in rotor coil; and, third, in certain brushless and stepper motors, the air gap between rotor (rotating part) and stator (fixed part), varies drastically with rotor rotation while the radial forces causing rotation vary with current. In all three causes, the current applied (confidential printer input) causes a proportional change in the size of an air column, resulting the production of sound waves with frequency components originating from motor hum, stator hum, and coil hum. Faruque et al.~\cite{FaruqueAcoustic3DPrinter} exploited this sound to propose the first attack against 3D printers. Using similar tools as keyboard side-channel attacks, namely the use of signal frequency features and supervised learning, they could extract the 3D printer style files corresponding to various objects with a recovery rate of 78\% in FDM printers. This approach of exploiting motor acoustics to infer inputs applies to all 3D printers based on motors including FDM, laser sintering, and laser sintering. Note that unlike the sound of a key (on a keyboard), the sound of a 3d printed object does not have a fixed frequency fingerprint---motor, stator, and coil hum frequencies change based on current applied. For this reason, using MFCC to extract the frequency component is not the best approach. In follow up work, Chhetri et al.~\cite{ChhetriConfidentiality} replaced MFCC with MODWT (Maximal Overlap Discrete Wavelet Transform) to capture a better fingerprint, increasing recovery rate from 78\% to 86\%.
Song et al. ~\cite{ccs16} use a smartphone stereo microphone and magnetometer together to better capture signal characteristics (Hojjati et al.~\cite{HojjatiSideChannelFactory} proposed the same for CNC milling machines). This approach has only incremental benefits since all motor inputs are already converted into acoustic sound due to magnetostriction, electrostriction, and radial forces on the rotor. Therefore combining acoustic with magnetic side-channels results in no fundamental improvement over audio side-channels.  
A number of works leverage acoustic side-channels to defend 3D printers. KCad~\cite{chhetri2016kcad} were the first to observe that integrity compromising attacks---false inputs in STereoLithograhy (STL) files that encode the CAD model), the GCodes, or firmware compromise---necessarily lead to acoustic emissions. They successfully isolated 3D integrity compromising attacks through supervised models.  Bayens et al.~\cite{usenixsec17} leveraged acoustic and other spatial layers emanations to verify the unseen internal fill structure present in 3D printed objects. They used microphones to record the sounds leaking from printer mounts and housing and trained an audio classifier to recover GCodes using peak frequency and its temporal location within recorded acoustic data. Their defense can verify 40--60\% of fill-pattern modification attacks. Belikovetsky et al.~\cite{belikovetsky2018digital} build on both the above approaches, to extend the defense coverage to 100\% of fill-modification attacks using  a PCA over the spectrogram of recorded sound.

{\bf Display hum:} The instantaneous power consumption of a display unit is a function of the screen content (processed in the raster sequence). This creates variations in the power supply causing  power-circuit components to vibrate due to electrostriction resulting in a power-acoustic transducer. This property generalises well beyond display circuits to all digital circuits where current varies as a function of the workload. Synesthesia~\cite{Synesthesia_2019} developed a passive ASC attack that leverages power-acoustic transduction to extract images from the audio traces of the display power supply, captured by a microphone and accessed by a remote attacker over a VoIP channel. However, they use specialist equipment (a large parabolic signal collector).

{\bf Fan hum:} A simple power-acoustic transduction occurs when heat triggers system cooling. Islam et al. ~\cite{macs18} analyse fan noise to determine power consumption thus developing a timing power attack rooted in acoustic signal analysis.

\subsection{Physical-key Leakage} 
Pin tumbler locks are widely used to secure homes and office spaces around the world. Recent work has developed methods to clone physical keys from the sounds emitted when a key is inserted. Ramesh et al.~\cite{ramesh2020listen} proposed SpiKey, which exploits the fact that each pin in  the tumbler makes a unique sound when depressed (just like a keyboard key). In follow up work, Ramesh et al.~\cite{usenixsec21} combined the acoustic signal with visual information to achieve a key recovery rate of up to 75\%.

\subsection{Acoustic Cryptanalysis} 

Genkin et al.~\cite{GenkinRSAAcoustic} introduced a passive acoustic cryptanalysis attack to extract full 4096-bit RSA keys with using the sound generated by the computer during the decryption of some ciphertexts. 
Using a phone or a sensitive microphone to record the sounds, the processed signals were then computed through a designed modular exponentiation which was based on the mathematical analysis of GnuPG (GNU Privacy Guard). 
Although this work has shown that different RSA keys induce different sound patterns that can be used to attack the keys, it was still not clear how to extract individual key bits. To address this issue, Genkin et al.~\cite{Genkin-2017} further expanded ~\cite{GenkinRSAAcoustic}. The main improvement of the key extraction is the time decision computation when performing the additional multiplication for every key bit. Compared to the previous version, this work built more detailed experiments to analyze the relevant code of GnuPG and experimentally showed that this acoustic key distinguishability is also possible on other ciphers, such as AES and DES, and other versions of GnuPG. 

\subsection{DNA Synthesis}
Faezi et al.~\cite{ndss19} proposed the first 
ASC attack on DNA synthesizer, where compromising confidentiality will leak valuable information on nucleotide sequences. 
Two sound sources were leveraged: 1) the unstable noise radiation 
caused by vibration when the
DNA synthesizer transports materials through the pipeline, 
2) the audible click produced by the DNA synthesizer when it opens and closes the flow of material. In the threat model, the DNA synthesizer can be connected to computers, external drives, and Ethernet cables, and it is impossible to tamper with the machine or access the output DNA sequence. The attacker must place at least one microphone to the DNA synthesizer within close physical proximity, which is a passive but non-invasive ASC.


\section{
Countermeasures}
\label{sec:Countermeasures}

To analyse countermeasures against ASC in a structured way, we use a three-dimensional framework, namely \textit{Impediment}, \textit{Interference}, and \textit{Obfuscation}. They represent three different defense principles respectively: preventing access to the ASCs, interfering with the observed signals, and obfuscating the original sound pattern with noise. We summarise these countermeasures in Table~\ref{countermeasure}, and note whether each of them was evaluated empirically.

\begin{table*}[htbp]
    \footnotesize
    \centering
    \vspace{-1.0cm}
    \caption{Acoustic side channels: Countermeasures}
    \label{countermeasure}
    \resizebox{\textwidth}{!}{
    \begin{tabular}{c|ccc|cccccc|c}
    \toprule

    \multicolumn{1}{c}{ } &  \multicolumn{9}{|c|}{\textbf{Countermeasures}} & \multicolumn{1}{c}{ } \\
\cmidrule(l){2-10}

    \multicolumn{1}{c|}{ } &  \multicolumn{3}{c}{\textbf{Principles }} &  \multicolumn{6}{|c|}{\textbf{Techniques} } & \multicolumn{1}{c}{ } \\
\cmidrule(l){2-10}

    {\textbf{Acoustic side channels}}  &   {\textbf{Im}} &  {\textbf{In}} &  {\textbf{Ob}} & {\textbf{\makecell{Acoustic \\ shielding}}} &{\textbf{\makecell{Stricter \\ access control}}} &  {\textbf{Alert}} &  {\textbf{\makecell{Add \\ noise}}} & {\textbf{\makecell{Randomization}}} &{\textbf{\makecell{Other techniques}}} & {\textbf{Evaluation}} \\ \midrule

     Asonov'04~\cite{asonov2004keyboard}&  \textcolor{blue}{\checkmark}   &  &  &  \textcolor{blue}{\checkmark} & &  &  &   & \makecell {Place the keys not in one plate}  & \textcolor{blue}{\textcolor{blue}{\checkmark} }  \\      \midrule

     Zarandy'20~\cite{zarandy2020hey}& \textcolor{blue}{\textcolor{blue}{\checkmark} }   &  \textcolor{blue}{\checkmark}   & & \textcolor{blue}{\checkmark}  &   &   &  \textcolor{blue}{\checkmark} & &   \makecell{Use phone cases or screen protectors}  & \textcolor{red}{\textcolor{red}{\xmark}} \\
        \midrule

   Backes'10~\cite {BackesAcousticPrinters}  & \textcolor{blue}{\checkmark}   &  &  & \textcolor{blue}{\checkmark} & \textcolor{blue}{\checkmark} & &  &  &    Longer distance & \textcolor{blue}{\checkmark}  \\ 
       \midrule

    Faruque'16~\cite{FaruqueAcoustic3DPrinter}&  \textcolor{blue}{\checkmark}   &  & \textcolor{blue}{\checkmark}   & \textcolor{blue}{\checkmark} &  &  & & \textcolor{blue}{\checkmark} &  \makecell{  Make the motor loads similar} & \textcolor{red}{\textcolor{red}{\xmark}} \\           \midrule
    
    Hojjati'16~\cite{HojjatiSideChannelFactory}& \textcolor{blue}{\checkmark}   & \textcolor{blue}{\checkmark}  &  & \textcolor{blue}{\checkmark} &  & & \textcolor{blue}{\checkmark} &   & \makecell{ Enlarge machines’s enclosures} & $\sqrt{}\mkern-9mu{\smallsetminus}$ \\          \midrule

   Keynergy'21~\cite{usenixsec21}&  \textcolor{blue}{\checkmark}   & \textcolor{blue}{\checkmark}  &  &  \textcolor{blue}{\checkmark} &  & & \textcolor{blue}{\checkmark} & &   \makecell { } & \textcolor{red}{\xmark} 
  \\             \midrule

   PIN Skimmer'13~\cite{Simon:2013:PSI:2516760.2516770} &  \textcolor{blue}{\checkmark}   &  &    &  & \textcolor{blue}{\checkmark}   & \textcolor{blue}{\checkmark}  & 
 & &  \makecell { } & \textcolor{red}{\xmark} \\ 
          \midrule
     
    Narain'14~\cite{SashankMicandGyro} & \textcolor{blue}{\checkmark}   &  &  &  &  \textcolor{blue}{\checkmark}  &  & &   & \makecell {Reduce sampling rate of the sensors } & \textcolor{red}{\xmark} \\ 
          \midrule

  SonarSnoop'18~\cite{sonarsnoop}  & \textcolor{blue}{\checkmark}  &  \textcolor{blue}{\checkmark}   &  & &   &  \textcolor{blue}{\checkmark} & \textcolor{blue}{\checkmark}  & &  \makecell{Disable the sound system; modify sensor design} & \textcolor{red}{\xmark} \\ 
        \midrule

   \makecell {PatternListener'18~\cite{zhou2018patternlistener} \\PatternListener+'19~\cite{sonarsnoop_similar}} &  \textcolor{blue}{\checkmark}   &  &  \textcolor{blue}{\checkmark}  &  & \textcolor{blue}{\checkmark} &  \textcolor{blue}{\checkmark} & & \textcolor{blue}{\checkmark} &  \makecell { Limit the frequency range \\ of the speaker and mic} & \textcolor{red}{\xmark} \\ 
         \midrule

     Shumailov'19~\cite{shumailov2019hearing}& \textcolor{blue}{\checkmark}  &  \textcolor{blue}{\checkmark}   &  & &  &   \textcolor{blue}{\checkmark} &  &  & \makecell {Inject fake taps; introduce timing jitter } & \textcolor{red}{\xmark} \\
        \midrule

  \makecell{Synesthesia'19~\cite{Synesthesia_2019}}  & \textcolor{blue}{\checkmark}   &  & \textcolor{blue}{\checkmark}   &  \textcolor{blue}{\checkmark} &  & &  &  &   \makecell{Make variations on software mitigations } & \textcolor{red}{\xmark} \\       \midrule

  Genkin'17~\cite{Genkin-2017} & \textcolor{blue}{\checkmark}   &   \textcolor{blue}{\checkmark}  &  \textcolor{blue}{\checkmark}  & \textcolor{blue}{\checkmark} &  & & \textcolor{blue}{\checkmark} &  \textcolor{blue}{\checkmark} &  \makecell{ } & \textcolor{red}{\xmark}  \\          \midrule

   KeyListener'19~\cite{yu2019indirect}  &  \textcolor{blue}{\checkmark}   &  &  \textcolor{blue}{\checkmark}   &  &  \textcolor{blue}{\checkmark} & &  & \textcolor{blue}{\checkmark}  & \makecell{ } & \textcolor{red}{\xmark} \\ 
          \midrule

    Oligo-Snoop'19~\cite{ndss19} & \textcolor{blue}{\checkmark}   &  \textcolor{blue}{\checkmark}  &   \textcolor{blue}{\checkmark} &  &  \textcolor{blue}{\checkmark}  & & \textcolor{blue}{\checkmark} &   \textcolor{blue}{\checkmark} & \makecell{ } & \textcolor{red}{\xmark}  \\     
     \midrule

  \makecell{Zhuang'05~\cite{zhuang2005keyboard}}& \textcolor{blue}{\checkmark} &   \textcolor{blue}{\checkmark}   &   & \textcolor{blue}{\checkmark} & &   & \textcolor{blue}{\checkmark} &   &  \makecell{}  & \textcolor{red}{\xmark}  \\            \midrule

  \makecell{Anand'16~\cite{fc16}} &  &   \textcolor{blue}{\checkmark}   &   &  &  & &  \textcolor{blue}{\checkmark} &    & \makecell { }& \textcolor{red}{\xmark} \\
     \midrule
    
   Skype \& Type'17~\cite{DBLP:journals/corr/CompagnoCLT16}  &  &   \textcolor{blue}{\checkmark}   &   &  & &   & \textcolor{blue}{\checkmark} &    & \makecell {Perform a short random transformation} & $\sqrt{}\mkern-9mu{\smallsetminus}$  \\
     \midrule

   \makecell{Anand'18~\cite{codaspy18}}  & &   \textcolor{blue}{\checkmark}   &   &  & &  &  \textcolor{blue}{\checkmark} &    &  \makecell{ } & \textcolor{blue}{\checkmark}  \\
     \midrule

   VoIPLoc'21~\cite{voiploc:wisec21} &  &   \textcolor{blue}{\checkmark}   &  \textcolor{blue}{\checkmark}  &   &  &  & &    & \makecell{Use acoustic jitter and network jitter}& \textcolor{red}{\xmark}  \\ 
    \midrule

     Song'16~\cite{ccs16} & \textcolor{blue}{\checkmark}  &   \textcolor{blue}{\checkmark}   &  \textcolor{blue}{\checkmark}    &  \textcolor{blue}{\checkmark} &      &   & \textcolor{blue}{\checkmark} & \textcolor{blue}{\checkmark} & \makecell{Inject additional dummy tasks }& \textcolor{red}{\xmark}  \\ 
    
    \bottomrule
    \end{tabular}}

 \begin{tablenotes}
     \item[1] 
Im:Impediment, In:Interference, Ob:Obfuscation, $\sqrt{}\mkern-9mu{\smallsetminus}$:partially evaluated. 
     \end{tablenotes}
    \vspace{-0.3cm}
\end{table*}

\subsection{Impediment}
Considering that getting access to target devices/systems or collecting useful acoustic signals is a necessary precondition for ASC attacks, to stop attackers from acquiring such acoustics, i.e. Impediment, is naturally an intuitive defense. Approaches include noise-dampening material or blocking the malicious application before access.

Asonov et al.~\cite{asonov2004keyboard} explore impediment defenses based on keyboard structure. They observed that keys located at different positions on a single mechanical plate will produce unique acoustic fingerprints, like tapping a drum in different places. They suggested developing {\em silent} keyboards with multiple sound-dampening plates and locating keys in acoustically equivalent locations to mitigate the attack. Zhuang et al.~\cite{zhuang2005keyboard} and Zarandy et al.~\cite{zarandy2020hey} also discussed these ideas and claimed that for mechanical keyboard emanations, the use of a silent keyboard is not an effective countermeasure, as the signal is still above the noise floor, unless each key is mounted on a separate plate. Zarandy et al.~\cite{zarandy2020hey} also mentioned that using phone cases or screen protectors may provide some measure of protection against acoustic side-channel snooping.

In the case of 3D printers and physical locks (both low-frequency ASC), noise reduction is a direct and effective measure. Regarding countermeasures against ASC attacks on printers, Backes et al.~\cite {BackesAcousticPrinters} tested the effectiveness of using acoustic shielding foam, placing the microphone at a larger distance, and placing the printer in another room. They found that ensuring the absence of sound collections in the printer's room is sufficient to resist most eavesdropping. A similar countermeasure was also considered in DNA synthesizer defense by Faezi et al.~\cite{ndss19}---prevent unauthorized person from entering the room. Faruque et al.~\cite{FaruqueAcoustic3DPrinter} and Song et al. \cite{ccs16} also suggested that shielding the 3D printer with a sound-proofing material can be considered as a countermeasure. Hojjati et al.~\cite{HojjatiSideChannelFactory } recommended improving shield motors, such as using composites to cover the stepper motors in manufacturing equipment, can help protect it from broadcasting sensitive information to an adversary. They also stated that enlarging the machines’ enclosures could help since magnetometer readings drop off with the cube of the distance from the source. 
In the case of physical keys, Ramesh et al.~\cite{usenixsec21} suggested modifying the lock design, such as making the key with noise-reducing material and removing the vulnerable key.

Early approaches to implementing the impediment have been crude---both these works suggest notifying users of the existence of side channels---in effect, asking the user to solve the sensor deadlock problem. To impede PIN inference attacks, Simon et al. ~\cite{Simon:2013:PSI:2516760.2516770} suggested using activity detection components at the OS level. When an activity is used to collect sensitive information from users, the component informs the OS and the OS will deny access to shared resources from other applications. Narain et al.~\cite{SashankMicandGyro} suggested blocking sensors in a mutually exclusive manner when a sensitive app runs. Cheng et al.~\cite{sonarsnoop} also proposed similar countermeasures to disable the sound system or notify users of a present sound signal in the high frequency range during sensitive operations to deal with gesture unlocking attacks which actively emit sound signals and use echoes to attack. Zhou et al.~\cite{zhou2018patternlistener, sonarsnoop_similar} discussed preventing the microphone from being used in the background and limiting the frequency range of the speaker and microphone. However, all these works fail to discuss how to deal with deadlocks that will naturally arise such as when app A has locked the accelerometer and waiting for the camera and app B does the same in reverse order. Another defense proposed by~\cite{sonarsnoop} is to modify sensor design to limit the supported frequency range, but this is challenging because deciding the threshold for cutoff is hard. A third approach as Zhou et al.~\cite{zhou2018patternlistener, sonarsnoop_similar}, Yu et al.~\cite{yu2019indirect} and Shumailov et al.~\cite{shumailov2019hearing} proposed is to notify the user and let them deal with it by disabling sound and/or sensors except touch screen during sensitive operations, this also seems inappropriate indicating that there is much further work to be done in impediment-based access control research.
For attacks of cryptographic key leaking and desktop display leaking, Genkin et al. ~\cite{Genkin-2017,Synesthesia_2019} propose acoustic shielding, however, this does not sit well with the need for air circulation to cool the heat. 

\subsection{Interference}
The working principle of interference defences is to drive the signal features the attack relies upon to well under the noise floor. 

The ASC attack for keyboard input has reached a certain degree of accuracy---attackers are exploring different advanced signal processing and classification algorithms to continuously improve the effectiveness of the attack, therefore disrupting the feature construction and classification process is a basic way for defenders.  
Zhuang et al.~\cite{zhuang2005keyboard} pointed out that quieter keyboards (Impediment) are useless. They believe that the ASC attack can be resisted by reducing the quality of the sound signal that the attacker may obtain, that is, increasing the noise. However, noise may also be separated, especially when faced with a microphone array attack, which records and distinguishes multiple channels of sound based on the location of the sound source. When an attacker is able to collect more data, this defense may also be ineffective. A smarter way proposed to add noise is to add a short noise window at each predicted peak, which may be more acceptable to users than continuous noise shielding. 
Anand et al.~\cite{fc16} proposed a 
defense mechanism against keyboard attacks which had good performance in the face of geometric measurement, feature classification, and other attack methods. The specific measure is to use 
background sounds to cover up the audio leakage.

The same is true for defense against remote attacks via VoIP. Compagno et al.~\cite{DBLP:journals/corr/CompagnoCLT16} proposed to perform a short random transformation of the sound when a keystroke is detected. The intuitive method is to apply a random multi-band equalizer on multiple small frequency bands of the frequency spectrum or mix the victim’s microphone with a masking signal to prevent remote attacks. Anand et al.~\cite{codaspy18} also believed that a noisy defense mechanism is feasible by generating a masking signal with speakers at the victim’s end, and those strategies were experimentally proved to be effective in protecting victims’ important information. 

Nagaraja et al.~\cite{voiploc:wisec21} also discussed a countermeasure for ASC attack on VoIP calls, while their target is to prevent location fingerprint leakage. Defenders may use acoustic jitter to damage the fingerprint information, such as using a constant amplitude signal at a room’s characteristic frequencies (50-2KHz) can cause a decrease in VoIPLoc’s performance. But it is hard to deploy because even small amounts of audible noise will negatively impact the voice quality, which is the first issue to be considered in VoIP.

In fact, this defense strategy of interfering with the original audio is effective for other different attack scenarios. Shumailov et al.~\cite{shumailov2019hearing} introduced timing jitter, or decoy tap sounds, into the microphone data stream to prevent attackers from reliably identifying tap locations when using virtual keyboards. As the taps themselves are pretty unnoticeable for humans, this should not disturb applications that run in the background. Another feasible countermeasure is to randomly play some distracting noises that are close to pressing when the virtual keyboard is used~\cite{zarandy2020hey}.
Cheng et al.~\cite{sonarsnoop} suggested a possible countermeasure against active ASC attacks is to block the propagation of inaudible sounds, such as generating inaudible noise to interfere, and when possible, refuse to receive low-frequency or high-frequency sound signals. 

The interference can still be applied to ASC attacks on 3D printers and physical key leaking. To protect 3D printing, Hojjati et al.~\cite{HojjatiSideChannelFactory} obfuscated the ASC emissions from manufacturing equipment by playing audio recordings of similar but flawed processes during production. Their experiments showed that such interference can make it harder for the attacker to separate the target audio stream from the others and reconstruct the object’s exact dimensions or process parameters. Song et al.~\cite{ccs16} also suggested introducing more interference during printing. Ramesh et al.~\cite{usenixsec21} thought that injecting noise to corrupt key insertion sounds is also a hopeful direction to improve security. Placing the machine in a noise environment has been discussed in Genkin et al.'s work~\cite{Genkin-2017}, but the noise is easily filtered by a high-pass filter due to the low frequency (below 10kHz) of the generated noise.
In the DNA synthesizer ASC scenario, Faezi et al.~\cite{ndss19} also suggested introducing additional noise by adding redundant physical components.

\subsection{Obfuscation}
One significant factor that causes keyboard acoustic attacks is that the keyboard always has a unified key layout, which makes an attacker easily infer the keys since the fixed location results in a distance pattern. Creating some similar noise with the target acoustics or randomizing the keys’ location (soft keyboard) can obfuscate the signals, thus hampering an adversary to infer the information correctly. 

This countermeasure is very useful and convenient to implement for the virtual keyboard on the touch screen, and it will not seriously affect the user experience. Compared with the physical keyboard, the layout of the touch screen virtual keyboard is easier to be customized, especially when inputting the PINs, the user's input habits can be temporarily ignored. For KeyListener, it needs prior knowledge of QWERTY keyboard layout to map localized keystroke positions to accurate characters. Therefore, Yu et al.~\cite{yu2019indirect} proposed that generating a random layout of the QWERTY keyboard is an effective way to resist touchscreen keystroke eavesdropping attacks. For the on-screen gesture unlocking leakage, a similar defense is to randomize the layout of the pattern grid~\cite{sonarsnoop_similar}. 

In addition to changing the position of the keys, randomization also plays a role in the defense against other attacks, such as cryptographic key leaking. Genkin et al. pointed out that their attack aimed at cryptanalysis can be prevented by some algorithmic countermeasures, such as ciphertext normalization and randomization~\cite{Genkin-2017}.

As for computer screen leaking, attacks can be defended against by changing the screen content. Genkin et al.~\cite{Synesthesia_2019} proposed that a more promising approach is software mitigation. Specifically, these programs cover leaks by changing the content on the screen, such as font filtering. By changing the font, all letters on the screen project the same horizontal intensity, avoiding the loss of information within a single pixel line. They also proposed two ways of shielding (impediment) and masking (interference), but these countermeasures are more difficult to achieve.

In fact, the defense strategy of obfuscation is also to prevent an attacker from extracting reliable information with distinct distinguishing characteristics. Nagaraja et al.~\cite{voiploc:wisec21} proposed a similar strategy, which is to use network jitter to induce packet latencies encouraging standard codec implementations to drop packets containing reverberant components, thus preventing the sender from extracting a credible room fingerprint. Moreover, Obfuscation can also be used for 3D printer and DNA synthesizer attacks. Faruque et al.~\cite{FaruqueAcoustic3DPrinter} suggested that creating similar loads on each motor and incorporating random motor movements can obfuscate the acoustic emissions. Song et al.~\cite{ccs16} considered adopting dynamic printing configurations in the process of G-code generation and injecting additional dummy tasks (e.g. use random trajectories). Faezi et al.~\cite{ndss19} suggested that operators can randomly select redundant steps of varying time length prior to delivery or randomly select and execute steps unrelated to base delivery to obfuscate signals.


\section{Discussions}\label{sec:findings}

We draw a number of interesting observations, which either reflect the strengths and weaknesses of the state of the art, or shed light on promising future research directions.

\textbf{Ever expanding attack surfaces.} Early work largely concentrated on physical keyboard emanation, and therefore targeted devices were PCs, laptops, payment devices and the like. The range of attack surfaces has been significantly expanded to date, covering smartphones, LCD displays, motherboards, mechanical locks, specialised equipment such as 3D printers and DNA synthesizers, and even computer-human interactions. Particularly, smartphones and 3D printers have attracted considerate attention in recent years. 

Overall, keyboard emanations have been the most studied among the ASCs. The second most studied is touchscreen leaking; followed by 3D printer leaking. Those less-studied categories are likely to offer more opportunities for future research. 
Where else to look for new ASCs? New devices and equipment where noise and sound are emitted will deserve a look.

\textbf{Data analysis and machine learning.}
The power of data analysis is critical for ASCs, as it hinges on the capability of extracting signals from often noisy data. There is a clear trend that ASC research evolved from simpler machine learning methods (e.g. probabilistic neural network, k-nearest neighbors, support vector machines) to more sophisticated deep learning (like convolutional neural network and recurrent neural network). As machine learning advances, it helps advance side-channel research. 

However, it is unnecessary that the more sophisticated the machine learning methods, the better. The nature of signals and the features of datasets collected all play an important role in choosing appropriate analysis methods. For example, Gohr \cite{gohr2019improving} reported at CRYPTO'19 some impressive cryptanalysis results achieved by deep learning. However, Benamira et al \cite{benamira2021deeper} showed at Eurocypt'21 that, after stripping down Gohr's deep neural network to a bare minimum, they achieved a similar accuracy using simple standard machine learning tools. 

In cases where deep learning does outperform simple machine learning methods, the black-box nature of the former can cause interpretability issues. For example, it may be unclear why the deep learning method has worked. What is its weakness? And, how to improve it? For example, Benamira et al. \cite{benamira2021deeper} achieved a complete interpretability of their method and the decision process, whereas Gohr \cite{gohr2019improving} fared poorly in explainability. 

\textbf{More nuanced nature of ASCs.} 
Early ASCs were passive ones, but recently active ASCs emerged~\cite{zhou2018patternlistener,sonarsnoop,lu2019keylistener}.  
Active ASCs are intriguing, as they involve with both intentional and accidental elements. Although acoustic signals were intentionally introduced by an attacker in active attacks, the signal-responses from the victim unintentionally leak information. 

Overall, most ASCs identified to date are passive ones, and only a few are active ones. Research into active ASCs is an interesting direction for future research.

We would not be surprised if many real-world attacks in the future will exploit a combination of active and passive ASCs, or exploit a combination of acoustic and other side channels, or simply amplify an ASC with non-side-channel attacks or vice versa. Certainly, researchers with imagination and creativity will be able to discover exciting new attacks along these directions, and only the sky is the limit.

\textbf{Constructive applications of ASCs.} Most research in this area 
employed ASCs for offensive purposes only, and several exceptions such as \cite{usenixsec17,panda2020behavioral,chhetri2016kcad,belikovetsky2018digital} looked into constructive or defensive applications of ASCs. Panda et al \cite{panda2020behavioral} investigated both offensive and defensive aspects of ASCs, where they attempted PIN guessing via keyboard emanation, as well as user verification via keystroke dynamics, which is a known behavioural biometric.
The basic idea of using ASCs to build security defenses is that acoustic signals emitted by devices can also be considered a fingerprint of the system or the program and used to protect the identification systems. It can be used alone or in combination with other protection mechanisms. This can be an exciting and promising direction for future research.

\textbf{Imbalance in attack and defence research.}
The literature has put significant effort into discovering new ASCs and their exploitation, rather than investigating countermeasures to them. In fact, we could only name a small portion that covered and discussed countermeasures. For this very reason, Table~\ref{countermeasure} is significantly shorter than Table~\ref{tab1}.

\textbf{Inadequate evaluations of countermeasures.}
What is worse, among those investigating countermeasures, only a small portion attempted empirical evaluations. Most countermeasures proposed remain theoretical  
Practical implementations and empirical evaluations are often limited, if any. 

Clearly, countermeasure investigations, in particular their empirical evaluations, have been under-appreciated and inadequate. 
Countermeasures lag behind attacks, and this may well suggest that the former may be much harder to deliver than the latter. However, all these no doubt warrant fertile grounds for future research. 

\textbf{Research methodology}. Experimentation is an intrinsic element of ASC research. However, experimental details are often under-reported in the literature. Thus, reproducibility can be a significant challenge. 

Moreover, many studies were mostly controlled experiments, conducted in strict laboratory settings or similar environments. There was inadequate effort in considering or pursuing whether the results could be generalized to other settings, in particular to the naturalistic real-world setting. Still much effort is required to demonstrate the ecological validity of these ASC studies. 
 
In terms of rigor and validity, ASC experiments in general are far behind the area of keystroke dynamics.
Via a series of solid works including 
\cite{killourhy2009comparing,maxion2011making,maxion2020reproducibility,wetherell2023effect},  Maxion's team at Carnegie Mellon meticulously examined and explored keystroke dynamics, and they achieved a high standard for repeatable, reproducible, well-grounded and generalizable experiments in security research. 
There is much for ASC researchers to learn from them.

Common metrics, reusable high-quality datasets, and standardized experimental setups and procedures (e.g. as shared operational protocols for experiments) all help to improve reproducibility. They  
will enable direct comparisons of attack or countermeasure research conducted by different teams. These will improve the rigor, validity and scientific foundation of ASC research, and advance the state of the art in an efficient and cost-effective way. 

\textbf{Lack of human, social and economic perspectives.}
Only a few papers (e.g. \cite{shumailov2019hearing,fc16}) considered usability and human factors, although some ASC countermeasures may potentially impact many users. On the other hand, monetary and computational costs incurred by potential countermeasures are rarely considered.  

Side channels could be hugely serious, 
with a far-reaching social and economic impact at a large scale, e.g. multi-billion dollar consequences. For example, following the discovery of differential power analysis \cite{kocher1999differential}, smart cards had to be redesigned for banking and other stakeholders all over the world. The microarchitectural side-channels like Meltdown~\cite{lipp2020meltdown} and Spectre~\cite{kocher2020spectre} suggested a major revisit of CPU designs, too. ASCs do not appear to be as serious. 

However, how serious can and will ASCs be in the future? Some security economic analysis can be relevant and interesting. To have an answer, it is critical to understand the severity, practicality, and impact of the various acoustic side channels in the real world. Which acoustic side channels pose a real threat? Or, most of them will remain of academic interest only? There are many interesting open problems.

\section{Bridging Side Channels and Inverse Problems} 

In unclassified worlds, side channels are a young field, with a history of less than forty years. Inverse problems have been studied for more than a century. 
However, side channels and
inverse problems appear to be two fields that are completely
isolated from each other\footnote{Some analysis in this section were initially developed for \cite{bourquard2022differential}.}.

A problem is \emph{inverse} because it starts with the observable effects to calculate or infer the causes, such as determining causal factors and unknown parameters from a set of measurements of a system of interest. It is the inverse of a forward---or direct---(physical) problem, which starts with the causes and then deduces or calculates the effects, such as modelling a system from known parameters. 

The field of inverse problems has deep and historical roots in mathematics, pioneered by giants like Hermann Weyl and Jacques Hadamard \cite{Kirsch_book, weyl1911asymptotische, hadamard1923lectures}. 
The main source of inverse problems is science and engineering.
These problems have pushed not only the development of
mathematical theories and tools, but also scientific and
technological innovations in a wide range of disciplines, including
astronomy, geophysics, biology, medical imaging, optics, and
computer vision, among others. Classical achievements of inverse problems include 
computed tomography (CT) and magnetic resonance imaging (MRI), where the inverse Radon transform is foundational.

\subsection{Side Channels versus Inverse Problems}

In a side channel, information leaks accidentally via some medium or mechanism that was not designed or intended for communication. Often, a direct measurement of the output from a side channel does not immediately give away the information leaked. Instead, the direct output measurement is akin to metadata, from which attackers deduce the leaked information.  

Therefore, \textbf{every side channel implies or involves an inverse problem, but not vice versa.} 

In some instances, a side channel may involve a relatively straightforward inverse problem. For example, 
Kuhn demonstrated a classical optical side-channel, where the information displayed on a computer monitor could be reconstructed remotely by decoding the light scattered from the face or shirt of a user sitting in front of the computer \cite{kuhn2002optical}. A sophisticated attack was required to successfully exploit this side channel. However, its key insight was the fact that the whole screen information was available as a time-resolved signal, rather than solving a complex inverse problem.
On the other hand, not all inverse problems involved in side channels are straightforward to solve. For example, active acoustic side channels such as SonarSnoop \cite{sonarsnoop}, KeyListener \cite{lu2019keylistener}, and PatternListener \cite{zhou2018patternlistener} all involved a rather complex inverse problem.

\subsection{Potential Impact on Side Channels}

How do the fields of inverse problems and side channels inform each other? 
We believe that the problem-formalisation strategies, theoretical models, mathematical techniques, algorithms, and concepts developed in inverse problems have significant potential to benefit and inspire future research of side channels (including acoustic ones). 

\textbf{First, it helps to properly navigate between the languages used in both fields.} This will, for instance, help to identify similarities and differences, to clarify misconceptions, and to unify terminologies.  
For example, \emph{information}, which is the set of relevant parameters approximated by the solution to the inverse problem, conceptually differs from \emph{measurements}, which are the physically leaked raw-data input of the inverse problem and which can contain various amounts of useful information. 

In a unified language that is understandable to both communities, blocking a side-channel attack essentially amounts to making the corresponding inverse problem unsolvable, intractable, harder to model, or at least harder to compute efficiently. Accordingly, there are the following three scenarios where one could: (a) prove that the inverse problem becomes impossible to solve by getting rid of the information that is present in the measurements, in such a way that the analysed measurements contain nothing relevant; (b) make the inverse problem much harder to model mathematically or solve computationally; (c) get rid of the leakage (e.g. physically) so that there are no measurements to exploit whatsoever, regardless of whether the said measurements would have contained meaningful information or not. Adding random perturbations such as noise is an example of a classical mechanism that makes an inverse problem unsolvable or harder to model.  

\textbf{Second, the perspective of inverse problems offers a new lens for examining side channels}.
As first elaborated by Jacques Hadamard, a fundamental challenge in inverse problems is they are typically ill posed in terms of the solution’s \textit{existence}, \textit{uniqueness}, and \textit{stability}, whereas their corresponding forward problems may be well posed in all these regards \cite{Kirsch_book}. 
The stability property means that a solution depends continuously on the available measurements (i.e. the observed data). Accordingly, a problem lacks stability if adding or removing data implies a radically different solution. If a computed solution lacks stability, it will simply depart from the true solution. 

Some studies of side channels (e.g. \cite{sonarsnoop,sonarsnoop_j}) may amount to only proving the existence of a solution for the corresponding inverse problem, rather than investigating the two related properties, namely, uniqueness and stability. Therefore, looking into these other properties, as studied from the perspective of inverse problems, will likely give security researchers a new lens for examining side channels, as well as their countermeasures.  

For example, examining the stability property alone warrants interesting research to answer the following questions. How will the side channel be impacted if less, or more, measurement data are collected for experiments? How much measurement data is necessary for the side channel to be stable, in such a way that the retrieved information depends continuously on the data, as opposed to varying abruptly across nearly similar datasets? Could specific countermeasures, such as adding some type of physical disturbance or interference, influence the observed output from the side channel in such a way that stability decreases? Answers to these questions could allow better optimising side-channel countermeasures, accurately simulating their expected effect before implementing them (e.g. in the case of optical side channels as demonstrated in \cite{bourquard2022differential}), quantifying their efficiency, and providing a robust framework to compare them in a systematic and rigorous manner.

\textbf{Third, some theoretical results on inverse problems are relevant to side channels.} One such result is reconstruction guarantees for several types of problem structures, such as lower bounds on reconstruction errors (Cramér-Rao bounds \cite{ye2003cramer}). These reconstruction guarantees are often only tied to the forward model mapping the relationship between the information of interest and measurements, in the sense that they do not depend on any specific algorithm or solution used. Another useful result is the extent to which the recovery is affected by noise or other non-idealities \cite{bungert2020variational,aster2018parameter}---which amount to mitigating side-channel attacks in 
security and cryptanalysis. Such results could inform one on how to best characterise various side channels---including acoustic, EM, 
and optical ones---and how to best design and evaluate their countermeasures. In particular, the interference and obfuscation countermeasures elaborated in Section \ref{sec:Countermeasures} can substantially benefit from the perspective of inverse-problem research due to their operational nature, even though impediment and some elements of obfuscation countermeasures may be out of scope for inverse problems. 

To solve challenging inverse problems, mathematics has been applied to accurately describe the forward model as well as assumptions on the solution, if any. 
For instance, sound statistical modelling allows reducing the dimensionality of the parameter spaces and producing accurate solutions \cite{kaipio2006statistical,scales2001prior}, and specific algorithms also allow maximizing computational efficiency. These may prove inspiring for side channel research, too.

Finally, it will be intriguing to explore possible connections between the optimality\footnote{By optimality, we mean that the maximum amount of information that can in theory be leaked from a side channel is fully extracted.} of a side channel in a given scenario and the uniqueness and stability of the solution to the corresponding inverse problem. In some cases, it appears that the latter indeed implies an optimal side channel. However, in many other scenarios, whether such a connection holds or not has no straightforward answers. Instead, these will be interesting areas for future research.


\section{Conclusions}\label{sec:conclusion}

We have seen steady progress in ASC research in the past twenty years. Some creative or even surprising results have emerged, such as acoustic cryptanalysis 
\cite{GenkinRSAAcoustic},
keyboard emanation 
\cite{asonov2004keyboard} 
and Synesthesia \cite{Synesthesia_2019}, to name a few. 

We have laid down some foundations to clear conceptual chaos, and put together a framework to structure our collective understanding of existing ASCs and their countermeasures. We have also identified gaps in the research, which point to promising future directions. 

We hope this paper sounds the marching bugle, attracting ambitious and creative researchers to further grow the field of ASCs, where imagination can make a difference. 

Finally, we have made an attempt to bridge side channels and
inverse problems. In general, every side channel implies (or involves) an inverse problem, but not vice versa. Although it may be a small step forward at this
stage, it is perhaps the start of an aspiration that will grow in the
future. We believe that this bridge has the potential to foster cross-field collaboration and inspire several new research directions,  
for example, building a more rigorous and effective scientific
foundation for side channel research, and encouraging the
possibility for ideas and techniques originated in one field to
enjoy a wider applicability than was previously anticipated.

\ifCLASSOPTIONcompsoc
  \section*{Acknowledgments}
\else
  \section*{Acknowledgment}
\fi
We thank Ilia Shumailov for his contribution, and Roy Maxion (Carnegie Mellon) for discussing experimental methods. PW and HCG were supported in part by the Natural Science Foundation of China under Grant 61972306 and 
by SongShan Laboratory under Grant YYJC012022005. This work was conceived and led by JY.

\bibliographystyle{IEEEtranS}
\bibliography{main}

\end{document}